\documentclass[prb,preprint,showpacs,preprintnumbers,amsmath,amssymb]{revtex4}
\usepackage{bm}
\usepackage{epsfig}

\begin{document}

\title[Excitation of spin waves on a cylinder]{Excitation of spin waves on a cylindrical semiconductor heterostructure with Rashba spin-orbit interaction}
\author{P. Kleinert}
\address{Paul-Drude-Intitut f\"ur Festk\"orperelektronik,
Hausvogteiplatz 5-7, 10117 Berlin, Germany}
\date{\today}
\begin{abstract}
Elementary excitations in a paramagnetic semiconductor quantum well confined to a cylindrical surface are theoretically studied on the basis of coupled spin-charge drift-diffusion equations. The electric-field-mediated eigenmodes are optically excited by an oscillating interference pattern, which induces a current in the outer circuit. For a cylinder with a given radius, sharp resonances are predicted to occur in the steady-state current response, which are due to weakly damped spin remagnetization waves.
\end{abstract}

keywords: spin-orbit interaction, cylindrical surface, interference pattern

\maketitle

\section{Introduction}
The spin-orbit interaction (SOI) in paramagnetic semiconductors renders possible an efficient control of the magnetization through an external electric field. This mechanism has received great interest in the emerging field of spintronics and has stimulated basic research of a two-dimensional electron gas (2DEG) with Rashba and Dresselhaus SOI \cite{RMP_323}. Well investigated spin-related phenomena in such systems are, in particular, the spin accumulation \cite{Edelstein} and the extrinsic and intrinsic spin-Hall effect \cite{Hirsch,SCIE_1348}. In addition, the electric field drives a rotation of the magnetic moment in a plane defined by the surface normal and the electric field vector \cite{PRB_205327}. Under certain conditions, this spin rotation creates elementary excitations, which are called spin-remagnetization waves \cite{Bryksin_JETP_2005}. These new spin-related eigenmodes can be excited by illuminating the sample with laser beams that generate both an oscillating space-charge grating of photoelectrons and an associated oscillating internal electric field. By exploiting this excitation technique, spin-remagnetization waves are generated, when the spin and charge degrees of freedom are coupled to each other \cite{Bryksin_2007}. Both the experimental set up and the theoretical analysis profit from the rich physics devoted to the study of space-charge waves in crystals. In a sense, this traditional field can easily be extended by considering the spin degree of freedom due to SOI. The excitation of novel spin-charge coupled, field-mediated eigenmodes is likewise possible by generating a definite photo-injected spin or charge pattern. Two laser beams with parallel linear polarization produce a grating in the free carrier concentration. Alternatively, an oscillating spin polarization is obtained by combining cross-linearly polarized pump pulses. A speciality of the spin subsystem is the appearance of long-lived spin excitations that are due to an exact spin-rotation symmetry. This so-called persistent spin helix \cite{PRL_236601} gives rise to strong field-induced resonances that can be detected by optical means \cite{PRL_076604W}.

Pronounced spin phenomena are expected to appear not only in a planar 2DEG, but also in two-dimensional carrier systems that are confined on a curved surface. Moreover, novel spin effects are predicted to occur because the curvature of the surface, in which the 2DEG resides, gives rise to additional contributions to the SOI \cite{PRB_085330}. In this paper, we focus on electric-field mediated spin excitations on a cylindrical 2DEG with Rashba SOI and short-range elastic scattering on impurities. For such a system, a specific long-lived spin excitation exists \cite{Trushin_2008}, which is studied under the condition that a constant electric field is applied along the cylinder axis. By taking into account the spin-charge coupling, the resonant excitation of this mode is demonstrated by generating an oscillating interference pattern. In the induced steady state current, pronounced resonances appear, when the radius $R$ of the cylinder matches the quantity $\hbar/(2m^{*}\gamma_1)$, where $\gamma_1$ denotes the Rashba spin-orbit coupling constant. These resonances have no counterpart in a planar 2DEG.

\section{Model and basic equations}
Let us treat a 2DEG with SOI of the Rashba type, in which the electrons are elastically scattered on impurities. An electric field $E_0$ applied parallel to the cylinder axis gives rise to a current density $j_0=en_0\mu E_0$ (with $n_0$ denoting the carrier density and $\mu$ the drift mobility) and affects damped eigenmodes of the system that have a combined spin-charge character. To excite these modes, an oscillating interference pattern is provided by two laser beams. The oscillation is achieved by phase modulation of one of the illuminating laser beams with frequency $\Omega$ and amplitude $\Theta$. A prerequisite of our approach is the isotropy of the illumination around the cylinder. We focus on the nonlinear excitation regime, in which effects proportional to the squared contrast ratio $m$ have to be taken into account. Under illumination, photogeneration $g(z,t)$ of electrons occurs, which is periodic both in time $t$ and along the cylinder axis $z$. The generation rate is given by
\begin{equation}
g(z,t)=g_0\left[1+m\cos\left(K_gz+\Theta\cos(\Omega t) \right) \right],
\label{e1}
\end{equation}
where $g_0$ is proportional to the total incident light intensity. The wave vector $K_g$ of the interference pattern is expressed by the spatial period $\Lambda$ via $K_g=2\pi/\Lambda$. By varying the angle between the two pump beams, the grating period can be tuned for the resonant excitation of coupled spin-charge modes.

The generated nonuniform carrier concentration $n(z,t)$ along the cylinder axis induces an electric field modulation $\delta E(z,t)$ so that the total internal electric field is given by
\begin{equation}
E(z,t)=E_0+\delta E(z,t).
\end{equation}
The self-consistent calculation of both quantities $n$ and $E$ accounts for the Poisson equation and is based on the following set of coupled nonlinear equations \cite{Kukhtarev,Bryksin_Petrov}
\begin{equation}
\frac{\partial n}{\partial t}+\frac{n-n_0}{\tau}+\frac{\varepsilon}{4\pi e}
\frac{\partial^2 E}{\partial z\partial t}=g(z,t),
\label{chargekin1}
\end{equation}
\begin{equation}
\frac{\varepsilon}{4\pi}\frac{\partial E}{\partial t}+j(z,t)=I(t),
\end{equation}
where $I(t)$ denotes the total current density in the outer circuit. In these equations, $\tau$ is the elastic scattering time and $\varepsilon$ the static dielectric permittivity. The internal current density $j(z,t)$ along the cylinder axis has an additional contribution, which is due to SOI. We obtain
\begin{equation}
j(z,t)=eD\frac{\partial n}{\partial z}+en\mu E+e\gamma_1\rho_{\varphi},
\label{chargekin3}
\end{equation}
where the diffusion coefficient $D$ satisfies the Einstein relation $\mu=eD/k_BT$. The spin-induced charge current is described by the component $\rho_{\varphi}$ of the density matrix, whose spin-related elements are collected in the vector ${\bm\rho}=(\rho_{\varphi},\rho_z,\rho_r)$. In the limit of a planar 2DEG ($R\rightarrow\infty$), $\rho_r$ converts to the out-of-plane spin density component. To close the set of self-consistent nonlinear equations for the calculation of the current in the outer circuit, the coupled spin-charge drift-diffusion equations are necessary that has been derived recently \cite{Klein_Zylinder}. These equations are expressed in a vector form
\begin{eqnarray}
\frac{\partial {\bm \rho}}{\partial t}&&+(D{\bm \kappa}^2-i\mu{\bm E}\cdot{\bm\kappa}){\bm\rho}
+\widehat{\Gamma}{\bm \rho}-\frac{2m^{*}}{\hbar}{\bm\rho}\times{\bm P}\nonumber\\
&&-4i\frac{m^{*}\tau}{\hbar}\left({\bm\Lambda}\times{\bm\omega}({\bm\kappa}) \right)n
-\frac{n\tau}{D}\widehat{\Gamma} {\bm P}={\bm 0},
\label{spinkin}
\end{eqnarray}
in which the spatial dependence of ${\bm\rho}$ is accounted for in the Fourier space by the wave vector ${\bm\kappa}$. The first three terms on the left-hand side of Eq.~(\ref{spinkin}) have the conventional form, in which the spin-scattering matrix $\widehat{\Gamma}$ appears. The remaining contributions describe the influence of an electric field on the spin polarization and the coupling to the charge density $n$, which has its origin also in the spatial inhomogeneity (${\bm\kappa}\ne{\bm 0}$). The SOI enters this equation via the coupling vector
\begin{equation}
{\bm\omega}({\bm\kappa})=-\gamma_1(\cos(2\varphi)\kappa_z,(1-\hbar/(2m^{*}R\gamma_1))\kappa_{\varphi},
\sin(2\varphi)\kappa_z),
\end{equation}
and the diagonal matrix $\widehat{\Gamma}$, whose elements are given by
\begin{equation}
\frac{2}{\tau_z}=4DK^2,\quad \frac{2}{\tau_r}=4DK^2\left[1+\left(1-\frac{\hbar}{2m^{*}R\gamma_1} \right)^2\right],
\end{equation}
\begin{equation}
\frac{2}{\tau_{\varphi}}=4DK^2\left(1-\frac{\hbar}{2m^{*}R\gamma_1} \right)^2.
\label{eq9}
\end{equation}
The momentum $K$ has its origin in the SOI: $K=2m^{*}\gamma_1/\hbar$ and $\varphi$ denotes an angle around the cylinder. It is worth mentioning that the spin-scattering rate $1/\tau_{\varphi}$ vanishes, when the radius of the cylinder satisfies the condition $R=\hbar/2m^{*}\gamma_1$. The vector ${\bm P}$ in Eq.~(\ref{spinkin}) is expressed by ${\bm\Lambda} +2iD{\bm\omega}({\bm\kappa})$ with ${\bm\Lambda} =(-\gamma_1\mu E,0,0)$.

Without any optical modulation ($m=0$), there is no dependence on the spatial ($z$) and temporal ($t$) variable. In this case, the analytic solution of Eqs.~(\ref{e1}) to (\ref{spinkin}) is easily obtained and has the form
\begin{equation}
n^{(0)}=n_0+g_0\tau,\quad E^{(0)}=E_0,
\end{equation}
\begin{equation}
I^{(0)}=en^{(0)}\mu E_0+e\gamma_1\rho_{\varphi}^{(0)},
\end{equation}
\begin{equation}
\rho_{\varphi}^{(0)}=-\frac{\gamma_1\tau}{D}\mu E_0n^{(0)},
\end{equation}
together with $\rho_r^{(0)}=\rho_z^{(0)}=0$. An electric-field induced spin polarization along the cylinder axis emerges, when initially a spin polarization $\rho_{0r}$ normal to the surface is provided. Under this condition, all components of the spin-density matrix are nonzero and we have
\begin{equation}
\rho_r^{(0)}=\frac{\rho_{r0}}{2/\tau_r+(\mu E_0)^2/D},
\end{equation}
\begin{equation}
\rho_z^{(0)}=\frac{\hbar}{2m^{*}\gamma_1}\frac{eE_0}{k_BT}\rho_r^{(0)}.
\end{equation}
These equations express the electric-field analogy of the Hanle effect \cite{Kalevich}. In the next Section, the steady-state current in the outer circuit is calculated, which results from the oscillating charge-density pattern.

\section{The steady-state current response}
Let us switch to the main task namely the treatment of effects, which are due to the optically generated interference pattern. The associated modulation of the charge density induces variations of the internal electric field ($Y=\delta E/E_0$), the charge current in the outer circuit ($f(t)=I(t)/I_0$, $I_0=en^{(0)}\mu E_0$), and the spin polarization (${\bm s}=\delta{\bm \rho}/n^{(0)}$). The corrections are calculated by an iteration of the nonlinear Eqs.~(\ref{chargekin1}) to (\ref{spinkin}) with respect to the small contrast ratio $m$. For a crystalline system without SOI, the solution was derived and discussed already previously \cite{Bryksin_Petrov}. Two space-charge waves were identified namely an oscillation of the free electron gas and a trap-recharging wave. Here, we are interested in modifications of the results, which are caused by SOI in a 2DEG on a cylindrical surface. The calculation is facilitated by exploiting the periodicity with respect to the space and time variables, which enables us to carry out a discrete Fourier transformation of Eqs.~(\ref{chargekin1}) to (\ref{spinkin}) via
\begin{equation}
Y(z,t)=\sum\limits_{p,l=-\infty}^{\infty}Y_{p,l}e^{ipZ+ilT},
\end{equation}
\begin{figure}
\centerline{\epsfig{file=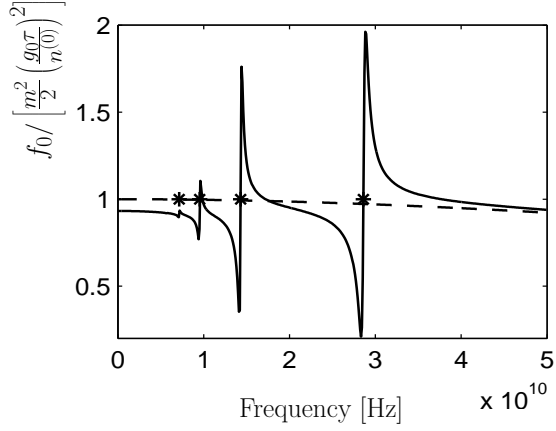,width=7.5cm}}
\caption{Induced relative stationary current as a function of frequency for $R=\hbar/(2m^{*}\gamma_1)$ (solid curve) and $R=5$~$\mu$m (dashed curve). Resonance frequencies $\Omega_r/(2\pi l)$ ($l=1,2,3,4$) are indicated by $(*)$. Parameters refer to the InAs based system: $\hbar\gamma_1=4\cdot 10^{-9}$~eVcm, $n^{(0)}=10^{15}$~cm$^{-3}$, $\Lambda=13.1$~$\mu$m, $m^{*}=0.023m_0$, $\varepsilon_0=15.15$, $E_0=2$~kV/cm, $\mu =1.75\cdot 10^{4}$~cm$^2$/Vs, $\tau=0.229\cdot 10^{-12}$~s, and $D=116.22$~cm$^2$/s.}
\label{Figure1}
\end{figure}
with $Z=K_gz$ and $T=\Omega t$. The SOI enters the approach only by the field-induced spin accumulation $\rho_{\varphi}$ in Eq.~(\ref{chargekin3}). This fact is in line with the observation that in a planar 2DEG with Rashba SOI, an in-plane electric field induces only a perpendicular in-plane spin polarization \cite{Edelstein}. The other two components of the spin-density matrix ${\bm\rho}$ vanish. Under the condition of weak SOI, we obtain from Eq.~(\ref{spinkin}) the solution
\begin{equation}
s_{\varphi;p,l}=\frac{ip\rho_{\varphi}^{(0)}/n^{(0)}}{il\Omega\tau_{E}-ip+\lambda p^2+2\tau_{E}/\tau_{\varphi}}
Y_{p,l},
\end{equation}
with $\lambda=DK_g/\mu E_0$ and $1/\tau_E=\mu E_0K_g$. By applying similar calculational steps as previously \cite{Bryksin_Petrov}, we obtain for the relative, optically generated steady-state current in the outer circuit the result
\begin{eqnarray}
f_0=&&\frac{m^2}{2}\left(\frac{g_0\tau}{n^{(0)}} \right)^2
\sum\limits_{l=-\infty}^{\infty}\frac{J_l^2(\Theta)}{|D_{et}|^2}
\biggl\{(1+l\overline{\omega}\lambda)\left|il\overline{\omega}+(\lambda-i)\frac{\tau_M}{\tau_E}+2\frac{\tau_M}{\tau_{\varphi}} \right|^2
\nonumber\\
&&+\kappa\frac{\rho_{\varphi}^{(0)}}{n^{(0)}}\frac{\tau_M}{\tau_E}
\left(l\overline{\omega}+(\lambda^2-1)\frac{\tau_M}{\tau_E}+2\lambda\frac{\tau_M}{\tau_{\varphi}} \right)\biggl\},
\end{eqnarray}
where $\tau_M=\varepsilon/(4\pi e\mu n^{(0)})$ denotes the Maxwellian relaxation time and $J_l(\Theta)$ is the Bessel function. In addition, we used the abbreviations $\overline{\omega}=\Omega\tau_M$ and $\kappa=\hbar\lambda/(2m^{*}D)(K/K_g)$. The optically generated constant current exhibits resonances due to field-dependent, coupled spin-charge excitations, the dispersion relation of which is determined from the zeros of the determinant
\begin{eqnarray}
D_{et}=&&l\overline{\omega} d(1+i\lambda)\left(il\overline{\omega}+(\lambda-i)\frac{\tau_M}{\tau_E}+2\frac{\tau_M}{\tau_{\varphi}} \right)
+\left(1+il\overline{\omega}\frac{\tau}{\tau_M} \right)\nonumber\\
&&\times\left[
(1+il\overline{\omega})\left(il\overline{\omega}+(\lambda-i)\frac{\tau_M}{\tau_E}+2\frac{\tau_M}{\tau_{\varphi}} \right)
+i\kappa\frac{\rho_{\varphi}^{(0)}}{n^{(0)}}\frac{\tau_M}{\tau_E}
\right].
\end{eqnarray}
The dimensionless parameter $d$ in this equation is given by $\mu E_0K_g\tau$. The cubic equation $D_{et}=0$ has three complex solutions $\overline{\omega}_i$, whose imaginary part determine the damping of the related eigenmode. Most interesting are nearly undamped excitations as their resonances are expected to be very pronounced. For the cylinder geometry there exists in fact a sharp resonance, when the radius $R$ of the cylinder matches the quantity $\hbar/2m^{*}\gamma_1$ so that the spin-scattering rate $1/\tau_{\varphi}$ vanishes as seen from Eq.~(\ref{eq9}). An example is shown in Fig. 1 by the solid line, which exhibits sharp resonances at frequencies $\Omega=\Omega_r/l$ ($l=1,2,3,4$) marked by stars. These pronounced resonances are oscillations of free carriers influenced by the SOI, whose eigenfrequencies are calculated from
\begin{equation}
\Omega_r=\mu E_0K_g\left(1+\frac{\tau}{2m^{*}D}\frac{\hbar^2K^2}{2m^{*}} \right).
\end{equation}
By detuning the resonance condition (dashed line in Fig.~1 calculated with $R=5$~$\mu$m), the markant features disappear. The parameters used in the calculation refer to the InAs based semiconductor system with relatively high mobility and drift velocity. Consequently, the resonance frequencies appear in the 10 GHz domain. Further theoretical and experimental studies of other material systems are necessary for meeting the conditions under which this effect can be detected in experiment.

In the absence of SOI, Eqs.~(\ref{chargekin1}) to (\ref{spinkin}) reduce to the known result \cite{Bryksin_Petrov}
\begin{equation}
f_0=\frac{1}{2}\left(\frac{mg_0}{n^{(0)}\tau_M} \right)^2
\sum\limits_{l=-\infty}^{\infty}J_l^2(\Theta)\frac{1+\lambda l\omega}
{|(\Omega-\Omega_1)(\Omega-\Omega_2)|^2},
\end{equation}
which describes space-charge waves with frequencies
\begin{equation}
\Omega_{1,2}=\frac{1}{2}\left(\frac{d}{\tau}+i\Gamma \right)\pm
\sqrt{\frac{1}{4}\left(\frac{d}{\tau}+i\Gamma \right)^2+\frac{1}{\tau\tau_M}}.
\end{equation}
The damping of these excitations is expressed by
\begin{equation}
\Gamma=DK_g^2+\frac{1}{\tau}+\frac{1}{\tau_M}.
\end{equation}
Recently, even the second harmonic generation of these space-charge waves was experimentally demonstrated \cite{PRB_085107} for photorefractive crystals.

\section{Summary}
The generation and manipulation of a spin polarization by an electrical field became a main subject in the emerging field of spintronics. In inhomogeneous systems, the spin and charge degrees of freedom are coupled to each other, which leads to spin-related eigenmodes, whose character is modified by an in-plane electric field. These so-called spin remagnetization waves can be excited by an optical grating technique \cite{Bryksin_2007}. Similar studies carried out on a 2DEG that is confined to a curved surface may reveal novel excitations that exclusively exist in a non-planar geometry. As an example, we treated a 2DEG on a cylinder and included in our model both the Rashba SOI and short-range elastic scattering on impurities. Eigenmodes of this system are excited by the optical generation of an oscillating interference pattern that matches the wavelength of the excitation. The irradiation of the sample induces a constant current in the outer circuit, which exhibits pronounced resonances for a special cylinder with a given radius ($R=\hbar/2m^{*}\gamma_1$). This robust field-dependent mode is weakly damped and does not exist in a planar 2DEG. The result confirms the expectation that novel spin effects may appear in systems with a non-trivial geometry or topology.


\begin{thebibliography}{10}

\bibitem{RMP_323}
I. Zutic, J. Fabian, and S.~D. Sarma, Rev. Mod. Phys. {\bf 76},  323  (2004).

\bibitem{Edelstein}
V.~M. Edelstein, Solid State Commun. {\bf 73},  233  (1990).

\bibitem{Hirsch}
J.~E. Hirsch, Phys. Rev. Lett. {\bf 83},  1834  (1999).

\bibitem{SCIE_1348}
S. Murakami, N. Nagaosa, and S.~C. Zhang, Science {\bf 301},  1348  (2003).

\bibitem{PRB_205327}
T. Damker, H. B\"ottger, and V.~V. Bryksin, Phys. Rev. B {\bf 69},  205327
  (2004).

\bibitem{Bryksin_JETP_2005}
V.~V. Bryksin, Sov. Phys. JETP {\bf 100},  314  (2005). [Zh. Eksp. Teor.
  Fiz. {\bf 127}, 335 (2005)].

\bibitem{Bryksin_2007}
V.~V. Bryksin and M.~P. Petrov, Sov. Phys. Solid State {\bf 49},  1906  (2007).
  [Fiz. Tverd. Tela {\bf 49}, 1818 (2007)].

\bibitem{PRL_236601}
B.~A. Bernevig, J. Orenstein, and S.~C. Zhang, Phys. Rev. Lett. {\bf 97},
  236601  (2006).

\bibitem{PRL_076604W}
C.~P. Weber, J. Orenstein, B.~A. Bernevig, S.~C. Zhang, J. Stephens, and D.~D.
  Awschalom, Phys. Rev. Lett. {\bf 98},  076604  (2007).

\bibitem{PRB_085330}
M.~V. Entin and L.~I. Magarill, Phys. Rev. B {\bf 64},  085330  (2001).

\bibitem{Trushin_2008}
M. Trushin and J. Schliemann, Physica E {\bf 40},  1446  (2008).

\bibitem{Kukhtarev}
N.~V. Kukhtarev, V.~B. Markov, S.~G. Odulov, M.~S. Soskin, and V.~L. Vinetskii,
  Ferroelectrics {\bf 22},  949  (1979).

\bibitem{Bryksin_Petrov}
V.~V. Bryksin, P. Kleinert, and M.~P. Petrov, Sov. Phys. Solid State {\bf 45},
  2044  (2003). [Fiz. Tverd. Tela {\bf 45}, 1946 (2003)].

\bibitem{Klein_Zylinder}
P. Kleinert, arXiv: 0808.0069 [cond-mat.mtrl-sci]  (2008).

\bibitem{Kalevich}
V.~K. Kalevich and V.~L. Korenev, JETP Lett. {\bf 52},  230  (1990). [Pis'ma Zh.
  Eksp. Teor. Fiz. {\bf 52}, 859 (1990)].

\bibitem{PRB_085107}
M.~P. Petrov, V. Bryksin, H. Vogt, F. Rahe, and E. Kr\"atzig, Phys. Rev. B {\bf
  66},  085107  (2002).

\end{thebibliography}
%
\section*{References}
%


\end{document}